\documentclass{article}

\usepackage{PRIMEarxiv}

\usepackage{listings}
\usepackage{caption}
\usepackage{setspace}
\usepackage{enumitem}
\usepackage{tcolorbox}
\usepackage[utf8]{inputenc} % allow utf-8 input
\usepackage[T1]{fontenc}    % use 8-bit T1 fonts
\usepackage{hyperref}       % hyperlinks
\usepackage{url}            % simple URL typesetting
\usepackage{booktabs}       % professional-quality tables
\usepackage{amsfonts}       % blackboard math symbols
\usepackage{nicefrac}       % compact symbols for 1/2, etc.
\usepackage{microtype}      % microtypography
\usepackage{lipsum}
\usepackage{fancyhdr}       % header
\usepackage{graphicx}       % graphics
\graphicspath{{media/}}     % organize your images and other figures under media/ folder

\title{ESBMC-Python: A Bounded Model Checker for Python Programs
}

\author{
  Bruno Farias \\
  University of Manchester \\
  Manchester, UK\\
  \texttt{bruno.farias@manchester.ac.uk} \\
   \And
  Rafael Menezes \\
  University of Manchester \\
  Manchester, UK\\
  \texttt{rafael.menezes@postgrad.manchester.ac.uk}
   \And
  Eddie B. de Lima Filho \\
  TPV Technology \\
  Manaus, Brazil \\
  \texttt{eddie.filho@tpv-tech.com} \\
   \And
  Youcheng Sun \\
  University of Manchester \\
  Manchester, UK \\
  \texttt{youcheng.sun@manchester.ac.uk}
   \And
  Lucas C. Cordeiro \\
  University of Manchester \\
  Manchester, UK \\
  \texttt{lucas.cordeiro@manchester.ac.uk}
}

\begin{document}
\maketitle

\begin{abstract}
This paper introduces a tool for verifying Python programs, which, using type annotation and front-end processing, can harness the capabilities of a bounded model-checking (BMC) pipeline. It transforms an input program into an abstract syntax tree to infer and add type information. Then, it translates Python expressions and statements into an intermediate representation. Finally, it converts this description into formulae evaluated with satisfiability modulo theories (SMT) solvers. The proposed approach was realized with the efficient SMT-based bounded model checker (ESBMC), which resulted in a tool called ESBMC-Python, the first BMC-based Python-code verifier. Experimental results, with a test suite specifically developed for this purpose, showed its effectiveness, where successful and failed tests were correctly evaluated. Moreover, it found a real problem in the Ethereum Consensus Specification.
\end{abstract}

% keywords can be removed

\onehalfspacing

\section{Introduction}
Python is an interpreted and multi-paradigm programming language to develop software systems, including general tasks, web applications, image processing, and artificial intelligence (AI)~\cite{van1995python}. Regarding the latter, the presence of Python code is particularly significant due to its extensive libraries, such as TensorFlow \cite{abadi2016tensorflow}, PyTorch \cite{paszke2019pytorch}, and Keras\cite{ketkar2017deep}. Indeed, its simple syntax, typing facilities, and resources made it popular, leading to its use in systems with critical security requirements. In contrast, its dynamic nature also hampers the development of static analyzers to ensure correctness.%, leading to bugs and weaknesses.

A technique that can be used to check Python programs, often employed for software verification, is bounded model checking (BMC) \cite{biere2021bounded}. Based on it, different languages can be tackled with specific tools or even front-ends for existing frameworks~\cite{monteiro2022model}. Moreover, the latter may harness the capacity of verification engines and then lead to more comprehensive and accurate results~\cite{menezes2022esbmc,song2022esbmc}.

%However, the BMC's potential for enhancing software quality in Python programs remains unexplored~\cite{phan2015concurrent}. Again, this is mainly due to its dynamic nature, which lacks explicit type information, unlike languages such as C \cite{madsen2015static}. Indeed, in Python, concrete-type information is assigned during program execution by its interpreter, which makes it harder for verification processes to evaluate code correctness, given that they rely on such knowledge. To address this issue, some studies have converted languages without explicit type information into C code, making model checking possible~\cite{monteiro2017bmclua}.

However, the BMC's potential for verifying Python programs remains unexplored~\cite{phan2015concurrent}. Again, it is mainly due to its dynamic nature, which lacks explicit type information, unlike languages such as C \cite{madsen2015static}. Indeed, in Python, concrete-type information is assigned during execution by its interpreter, which makes it harder for verifiers to evaluate correctness, given that they rely on such knowledge. To tackle this, some studies have converted languages without explicit type information into C code, making model checking possible~\cite{monteiro2017bmclua}.

Although this seems to lead to a dead-end, an aspect should be mentioned: the Python syntax allows annotation with type information on variables and functions. Consequently, this instrument, together with other resources, such as abstract syntax trees (ASTs) and satisfiability modulo theories (SMT) or Boolean satisfiability (SAT) solvers, could be used for reasoning about a program's states. 

In other words, type annotations in Python code could favor its analysis by a BMC tool, such as the efficient SMT-based bounded model checker (ESBMC) \cite{cordeiro2011smt}. This formal verifier has already been applied to many systems, including digital filters~\cite{AbreuGCFS16}, controllers~\cite{BessaICF14,ChavesIBCF19}, and unmanned aerial vehicles~\cite{ChavesBIFCF18}. Such a track record assures a distinct level to it, which, with new languages and features, can expand its applicability and provide a functional approach.

The last paragraphs outline the inspiration for the present work, which proposes a scheme to make BMC tools capable of processing Python code. It converts the latter into an AST structure, which is then type-annotated and formatted to provide a description suitable to a BMC pipeline. Aiming at evaluation, we have implemented this approach using ESBMC due to its mature and well-proved engine.

The proposed approach includes a front-end to generate ASTs from Python programs. These elements serve as interfaces between Python source code and the ESBMC's internal model-checking structure, thus translating a program into an intermediate representation (IR) that it can analyze. Subsequently, the ESBMC's back-end generates first-order logical formulae for a program's constraints and safety properties, aiming at formal verification. Ultimately, such formulae are submitted to an SMT solver to check for satisfiability.

The resulting tool was named ESBMC-Python and used to evaluate a benchmark suite. The latter is a collection of Python programs created to assess our tool and allow comparison with similar ones, which is another contribution of ours that can be used for evaluating Python verifiers. In this context, ESBMC-Python verified Python programs in a few milliseconds ($30$ ms on average), automatically identifying violations related to user-defined assertions and arithmetic and logical operations. We have also used ESBMC-Python to check the Ethereum consensus specification \cite{cassez2022formal}. As a result, it found an issue confirmed and fixed by the respective maintainers.

\section{Tool Description}
\label{sec:headings}

ESBMC receives source code as input and generates AST descriptions, which identify the different operations within a program and include relevant information, such as statement location and variable type and size. Next, program statements are translated into symbols and added to a structure representing a program's symbol table (ST) using the ESBMC's IR format (IRep). Indeed, the key aspect of using the BMC pipeline implemented in ESBMC is its ST, which must be the final result of a Python front-end.

Moreover, Python offers type annotation for variables, function parameters, and return values, which does not affect its runtime behavior or generate errors. Indeed, it should be used in static analysis to complement the information in a resulting AST description. Consequently, we have implemented ESBMC-Python using the libraries \textit{ast}~\cite{python_ast} and \textit{ast2json}~\cite{ast2json} together with ESBMC.

\subsection{The Python Verification Architecture}
\label{sec:arch}
\begin{figure}[htb]
\begin{center}
\includegraphics[width=10cm]{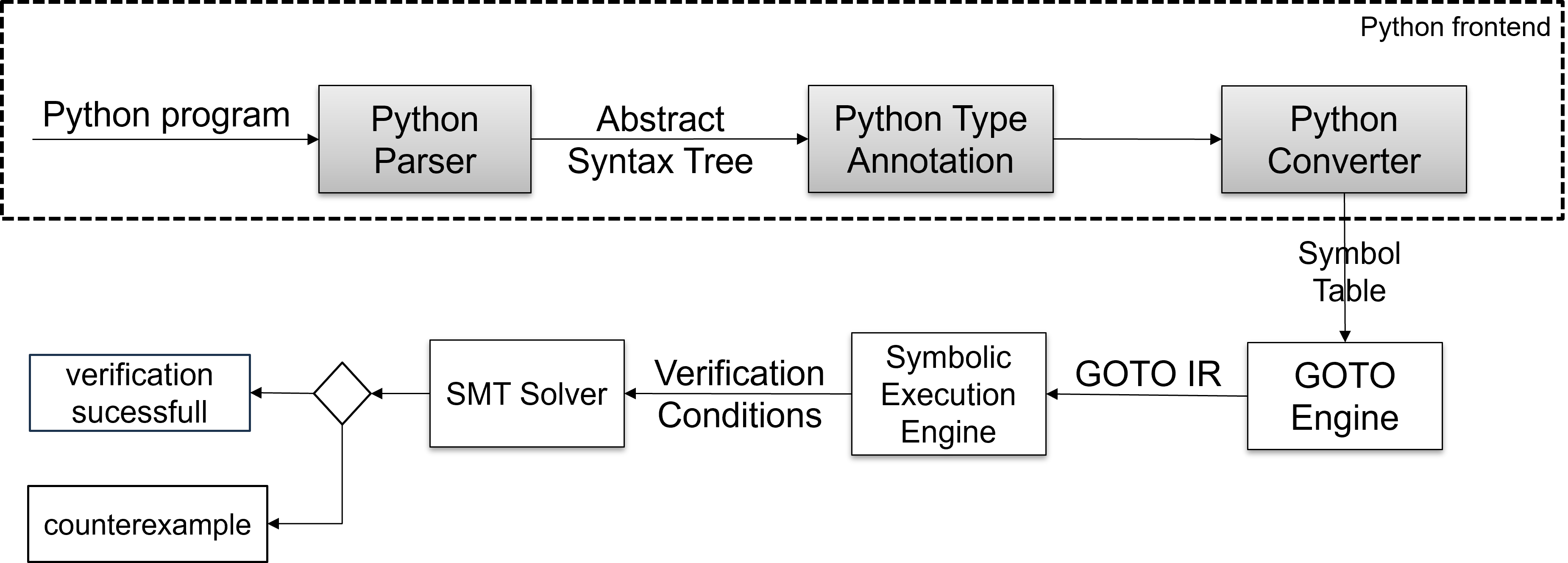}
\end{center}
\caption{ESBMC-Python architecture.}
% \vspace{-10.4mm}
\label{fig:workflow}
\end{figure}

The Python verification scheme proposed here results directly in a front-end for ESBMC, reusing its infrastructure and back-end components. This complete arrangement is called ESBMC-Python, whose architecture is shown in Fig.~\ref{fig:workflow}. The front-end includes code parsing, semantics analysis, and statement translation into an ESBMC's ST. The gray elements represent new front-end components, while the others are preexisting ones reused by ESBMC-Python.

\textbf{Python Parser}. The processing begins with {\it Python Parser}, which analyzes the input code structure and generates the corresponding AST, using the module \textit{ast}. Then, the respective output is passed to \textit{ast2json} for JSON conversion. 
Specifically, an instance of the Python interpreter is invoked to run a script that employs these libraries, ensuring that a program's behavior is correctly represented. Additionally, {\it Python parser} can handle user options from ESBMC to print AST content, i.e., {\it --parse-tree-too}, and isolate functions, i.e., {\it --function}. This last feature allows the verification of an entire file or its functions, which is useful to avoid the analyses of unsupported parts or reduce verification times. The output of this component is a file \textit{ast.json} containing the input program's structure.

\textbf{Python Type Annotation}. Next, {\it Python Type Annotation} adds type annotations to the output AST using type inference~\cite{duggan1996explaining}, i.e., it inserts new nodes with typing information for a program's variables. This task involves processing an AST and adding nodes \textit{AnnAssigned} containing the field \textit{id} with specific type information~\cite{python_ast}.

\textbf{Python Converter}. Ending the front-end processing, {\it Python converter} turns the definition of classes, methods, and functions into an ST in IRep. Specifically, it iterates over expressions in functions, conditional blocks, and loops, converting each operation with the available ESBMC's application programming interface (API).

Subsequently, the existing ESBMC's pipeline retrieves expressions from the resulting ST and converts them into GOTO language, which is considered another IR and represents its control flow graph (CFG). It transforms the program logic into a simplified representation based on assignments, conditional and unconditional branches, assumes, and assertions. Then, a symbolic execution interprets a bounded execution of the GOTO program, resulting in its static single assignment (SSA) trace \cite{cytron1991efficiently}. In SSA forms, all assignments over variables construct new symbols that can be combined using $\phi$-functions, ultimately generating Boolean formulae. Such elements represent a verification condition (VC) $C \land\neg P$ submitted to a solver, where $C$ means constraints and $P$ denotes a safety property. 

If a specific function $F$ is verified, using {\it –function}, {\it Python Converter} converts only $F$ and then adds its call, passing non-deterministic values as parameters. This process evaluates $F$ with all possible values of a given type, helping identify issues that often go unnoticed.

Note that ESBMC was not designed to handle object-oriented programming (OOP)\cite{cordeiro2011smt}. Therefore, we had to model OOP features %(e.g., inheritance, polymorphism, and method overloading) 
with structured programming. For instance, {\it Python Converter} resolves calls to overloaded or inherited methods by searching for their definitions in base classes, respecting class ordering in inheritance lists. Moreover, when class attributes are not defined for a given class, it looks for them in base classes from the respective ST.

ESBMC-Python supports built-in types, i.e., \textit{int}, \textit{float}, and \textit{boolean}, and basic structures. The latter include logical operations, comparisons, assignments, asserts, conditionals, loops, functions, module imports, classes, inheritance, and polymorphism. Finally, regarding verification properties, it can detect division by zero, arithmetic overflows, out-of-bounds array access, and user-defined assertions.

%--------------------------------------------------
\subsection{Illustrative Example}
%--------------------------------------------------

In this section, we explain tool usage for the program in Listing~\ref{lst:example}, which includes a recursive function for the factorial of an integer. Its input variable $n$ is initialized with a non-deterministic value at line $7$, whose range, with calls to \textit{ESBMC\_assume}, is constrained to $[1,5]$, at line $8$. Then, $factorial$ is called, at line $9$, and, finally, an assertion checks that its return can not be $120$, at line $10$. 

Listing~\ref{lst:ast} contains the AST in JSON format corresponding to the assignment to $n$ that occurs at line $7$ of Listing~\ref{lst:example}. Indeed, ESBMC-Python manipulates an AST by transforming simple assignments into annotated ones during its program parsing process explained in Section~\ref{sec:arch}, using {\it AnnAssign} nodes. In this case, the variable {\it result} is initialized with the value returned by {\it nondet\_int}.

% Moreover, its input parameter and return value are annotated, in line $1$, which also happens with the variable that stores the call result, in line $7$, and is necessary for ESBMC-Python. However, this call is followed by an assertion, where the result $20$ is checked instead of $24$, making this verification fail. In summary

ESBMC-Python verifies whether the negation of a property is satisfied, which becomes a satisfiability problem. Using SSA, it creates $C$ with assignments, interval restrictions, and the factorial computation itself, leading to $C = [n = nondet() \land n > 0 \land n < 6 \land result = \phi(n = 1 \rightarrow 1, n = 2 \rightarrow 2, n = 3 \rightarrow 6, n = 4 \rightarrow 24, n = 5 \rightarrow 120)]$, while $P$ uses line $10$, resulting in $P = [result \neq 120$]. Then, the corresponding VC is submitted to a solver, which tries to find a value combination that satisfies $C\land\neg P$.

This program can be verified by running the binary $esbmc$ \footnote{\url{https://github.com/esbmc/esbmc/releases/tag/v7.6.1}} with our Python front-end already integrated, passing its Python file name as a parameter. Assuming a file \textit{main.py}, our tool can be executed using the command:

\begin{verbatim}
  $ esbmc main.py --unwind 5
\end{verbatim}
where $-unwind$ limits the recursion depth during symbolic execution.
  
%\noindent The parameter {\it --unwind} limits the loop unfolding for symbolic execution.

Fig.~\ref{fig:esbmc_output} contains the verification output for the same program in Listing~\ref{lst:example}. As one may notice, the solver within the ESBMC's pipeline detects a value of $120$ when $n$ is equal to $5$.

% \begin{lstlisting}[numbers=left
%                   ,xleftmargin=4.3ex
%                   ,label={lst:example}
%                   ,language=Python
%                   , frame=single
%                   , basicstyle=\footnotesize\ttfamily
%                   ,caption={A Python program verifiable by ESBMC-Python.}]
% def factorial(n:int) -> int:
%     if n == 0 or n == 1:
%         return 1
%     else:
%         return n * factorial(n - 1)

% n:int = nondet_int()
% __ESBMC_assume(n > 0 and n < 6)
% result:int = factorial(n)
% assert(result != 120)
% \end{lstlisting}

% \begin{lstlisting}[numbers=left
%                   ,xleftmargin=4.3ex
%                   ,label={lst:ast}
%                   , frame=single
%                   , basicstyle=\footnotesize\ttfamily
%                   ,caption={AST in JSON for an annotated assignment.}]
% {
%   "_type": "AnnAssign",
%   "annotation": {
%       "_type": "Name",
%       "id": "int",
%   },
%   "target": {
%     "_type": "Name",
%     "id": "n"
%   },
%   "value": {
%     "_type": "Call",
%     "args": [],
%     "func": {
%       "_type": "Name",
%       "id": "nondet_int",
%     },
%   }
% }
% \end{lstlisting}

\vspace{1\baselineskip}

\begin{minipage}{0.44\textwidth}
\begin{lstlisting}[label={lst:example}, language=Python, caption={A Python program verifiable by ESBMC-Python.}]
def factorial(n:int) -> int:
  if n == 0 or n == 1:
    return 1
  else:
    return n * factorial(n - 1)

n:int = nondet_int()
__ESBMC_assume(n > 0 and n < 6)
result:int = factorial(n)
assert(result != 120)
\end{lstlisting}
\end{minipage}
%\hfill
\hspace{0.5cm}
\begin{minipage}{0.46\textwidth}
\begin{lstlisting}[label={lst:ast}, caption={AST in JSON for an annotated assignment.}]
{
  "_type": "AnnAssign",
  "annotation": {
      "_type": "Name",
      "id": "int",
  },
  "target": {
    "_type": "Name",
    "id": "n"
  },
  "value": {
    "_type": "Call",
    "args": [],
    "func": {
      "_type": "Name",
      "id": "nondet_int",
    },
  }
}
\end{lstlisting}
\end{minipage}

\begin{figure}[htb]
\begin{center}
\begin{minipage}{0.58\textwidth}
\begin{lstlisting}[language=bash, frame=single, basicstyle=\footnotesize\ttfamily]
ESBMC version 7.6.1 64-bit x86_64 linux
Parsing main.py
Converting
Generating GOTO Program
GOTO program creation time: 0.023s
...
Building error trace
[Counterexample]
State 1 file main.py line 7 column 0 thread 0
-------------------------------------------
n = 5 (00000000 00000000 00000000 00000101)

State 4  thread 0
-------------------------------------------
Violated property:
  assertion
  result != 120
  
VERIFICATION FAILED
\end{lstlisting}
\end{minipage}
\end{center}
\caption{ESBMC-Python output.}
\label{fig:esbmc_output}
\vspace{-5mm}
\end{figure}

\section{Experimental Evaluation}
\label{sec:exp}

Here, we present ESBMC-Python's verification results, which intend to answer two experimental questions (EQ):
%
%\begin{tcolorbox}
\begin{itemize}[leftmargin=*]
    \item {{\bf EQ1 (soundness)}} - can our approach report known wrong programs and preserve the correct ones, presenting soundness?
    \item {{\bf EQ2 (performance)}} - what are the time and memory performances associated to our approach?
\end{itemize}
%\end{tcolorbox}
\noindent In this context, soundness refers to the capacity to ensure that no correct program is considered wrong.
 
To answer these questions, we created a benchmark suite of $85$ programs\footnote{\url{https://github.com/esbmc/esbmc/tree/master/regression/python}}, each named for its target feature, split across $15$ categories. There are at least two tests: one with a failing assertion %, appended with ``fail'', 
and the other with one or more passing assertions. We also created extra elements for sensitive features such as imports and functions. %We will showcase a unique test result from each feature, highlighting the variety of features covered in this work. %The test suite, available in the ESBMC's repository, includes tests that explore specific feature of the Python language, demonstrating the tool's capability to handle different Python expressions. 

Moreover, we assess ESBMC-Python's performance in handling different Python expressions by measuring memory usage and verification times using the Linux \textit{time} tool. The obtained results report total computer processing unit (CPU) times, including user and system portions, thus gathering the real CPU occupation.

Our benchmark suite encompasses features usually found in real-world Python programs: arithmetic operations, conditionals, loops, user assertions, bit-wise operations, classes, objects, class attributes, instance attributes, inheritance, polymorphism, function definitions, function calls, recursive functions, module imports, non-determinism, and assume directives. This way, it should not be considered only suitable for simple validation or a set of toy examples, which often happens in initial studies~\cite{shu2019model}. Indeed, the absence of benchmarks for Python verification underscores its importance as a possible baseline for future investigations.

%We have written tests to manipulate variables across various program flows, with a subsequent user assertion checking for the expected values. ESBMC-Python can verify both programs with assertions that should fail, as well as those where all assertions hold true, showing its ability in verifying both property violations and adherence.

% Table \ref{tab:results} provides results for the created dataset, informing functionality, execution time, and outcome. Currently, ESBMC-Python takes, in average, around $30$ms for simple programs that thoroughly explore a functionality and use assertions to check the program's state. The latter occurs by verifying the value of variables altered during a program execution.

We checked our benchmark suite on a 64-bit Intel i$7$-$12700$H processor with $16$ GB of RAM, running Ubuntu $22.04$. Moreover, we used version $7.6.1$ of ESBMC, following the compilation instructions in its project documentation ~\footnote{\url{https://github.com/esbmc/esbmc/blob/master/BUILDING.md}}. Specifically regarding the ESBMC's verification pipeline, we employed version $3.2.3$ of the solver Boolector \cite{brummayer2009boolector}.

All verification processes were successful. This way, ESBMC-Python identified programs with property violations and validated the ones with correct behavior, which answers {\bf EQ1}. 

\begin{tcolorbox}
\textbf{EQ1}: ESBMC-Python only detected property violations for wrong program elements, which included types, conditionals, loops, functions, user assertions, and OOP aspects.
\end{tcolorbox}

% All tests with assertions that can be checked statically, i.e., without solver execution, were verified in around $25$\,ms, whereas those involving verification conditions took approximately $30$\,ms. This, in turn, answers {\bf EQ2}. Regarding memory usage, our tool consumed, on average, $19$MB to verify a program. Besides, among all the tests on our benchmark suite, the highest value was $27$MB. Indeed, such figures are considered low regarding modern personal computers.
Table \ref{tab:results} summarizes average results for memory usage and execution time, per test category. The highest and lowest average verification times were $49.1$ ms and $24.5$ ms, respectively, which, compared to what is obtained with BMC tools for similar programs in other languages, can be regarded as satisfactory \cite{kroening2014cbmc}. It also means that large project repositories or extensive program sets could be verified in relatively short periods, automatically \cite{de2023finding}. Regarding memory consumption, the highest and lowest amounts were $26.4$ MB and $14.5$ MB, respectively, which is also usual \cite{kroening2014cbmc}. %, compared with similar programs in other languages %tests with assertions that can be statically checked, i.e., without solver execution, consumed an average of $19$ MB. In contrast, those requiring solver invocation consumed around $15$ MB. 
Moreover, the highest memory usage for an isolated test was $27$ MB, which occurred in category {\it Classes}. It seems to be due to the representation of instance attributes and the necessary search for base classes, when inheritance is involved. Nevertheless, these figures are still considered low for modern personal computers. Finally, such results provide a summarized view of the ESBMC-Python's performance, %These results indicate that while solver invocation has a minimal impact on execution time, it does affect memory usage, 
thereby addressing \textbf{EQ2}. 

\begin{table}[htb]
    \centering
    \setlength{\abovecaptionskip}{5pt}
    \begin{tabular}{|c|c|c|c|}
        \hline
        Category & Test Cases & Mem. Usage & Exec. Time\\
        \hline
        Arith operations & 2 & 26.4 MB & 33.5 ms\\
        \hline
        Assignments & 5 & 18.5 MB & 38 ms\\
        \hline
        Assume & 4 & 16.5 MB & 28.2 ms\\
        \hline
        Binary operations & 2 & 20.5 MB & 29.5 ms\\
        \hline
        Binary types & 4 & 20.4 MB & 28.5 ms\\
        \hline
        Built-in functions & 7 & 19.9 MB & 28.1 ms\\
        \hline
        Classes & 9 & 19 MB & 27.1 ms\\
        \hline
        Conditionals & 4 & 17.8 MB & 25.5 ms\\
         \hline
        Functions & 11 & 21.8 MB & 30 ms\\
        \hline
        Imports & 8 & 15.3 MB & 49.1 ms\\
        \hline
        Logical operations & 6 & 20.4 MB & 24.5 ms\\
        \hline
        Loops & 10 & 20.7 MB & 35.4 ms\\
        \hline
        Non-determinism & 4 & 21.4 MB & 29.2 ms\\
        \hline
        Numeric types & 6 & 20.9 MB & 29.1 ms\\
        \hline
        Type annotation & 3 & 14.5 MB & 27.3 ms\\
        \hline
%        {\bf Total} & {\bf 85} & - & -\\
%        \hline
    \end{tabular}
    \caption{Results for ESBMC-Python regarding our test suite.}
    \label{tab:results}
%    \vspace{-8mm}
\end{table}

\begin{tcolorbox}
\textbf{EQ2}: ESBMC-Python presented execution time and memory consumption figures that are similar to what is noticed for BMC tools targeting other languages.%demonstrated robust performance, consistently verifying 
%verified programs with different expressions, statements, and features.
\end{tcolorbox}

%Currently, ESBMC-Python takes, in average, around $30$ms for simple programs that thoroughly explore a functionality and use assertions to check a program's state. The latter occurs by verifying the value of variables altered during a program execution.

As far as we know, only one tool is similar to ESBMC-Python: modeling, simulation, and verification (MSV)~\cite{shu2019model}. However, there is no test set, reproducible results, or repository for retrieving its source code (see Section \ref{rel-work}), which impedes a direct comparison.

%---------------------------------------
\subsection{Experimental Results for the Ethereum Consensus Specification}
%---------------------------------------

We also used ESBMC-Python to check the Ethereum blockchain consensus protocol, which comprises elements that control the Ethereum network's node inclusion, validation, and validator penalty processes. It is described with a set of markdown files, in a GitHub repository~\footnote{\url{https://github.com/ethereum/consensus-specs}}, containing functions that compose a reference API used to generate a Python library called \textit{eth2spec}.

All functions in this specification include parameters and return values with annotated typing. This way, one could submit \textit{eth2spec} for evaluation by ESBMC-Python. However, it also contains elements not initially handled by ESBMC-Python, which led to extensions for custom types such as \textit{uint64}, \textit{uint128}, and \textit{uint256}.

We used ESBMC-Python to verify Python files generated for \textit{eth2spec}, which involved testing each function individually (see Section~\ref{sec:arch}). Specifically, we employed the parameter {\it --function} followed by the name of the function to be verified, thus checking specific elements with non-determinism in their invocations.

As a result, our evaluation revealed a division-by-zero in function \textit{integer\_squareroot}. This error occurred because an unsigned integer overflowed to zero, after being incremented, and was later used as the denominator of a division operation. Indeed, division-by-zero events in blockchains can lead to service interruptions, compromising their availability and facilitating attacks.

%---------------------------------------
\section{Related Work}
\label{rel-work}
%---------------------------------------

Shu {\it et al.} \cite{shu2019model} proposed using the MSV language (MSVL) to describe and check Python programs with the MSV tool, utilizing rules to express Python's semantics in MSVL. %Once a program is encoded into MSVL, it can be verified with the MSV tool. 

Although this work proposes a technique for automatic verification, its examples and functionalities are still basic. Hence, we can state that the MSV tool can not verify complex Python programs as found in industrial applications. Moreover, it is not available for download, preventing its evaluation, comparison, and further development. We have indeed attempted to contact its authors by email, but we did not receive a response while writing this paper.

%-----------------------------------------------------
\section{Tool Availability}
\label{tool-avail}
%-----------------------------------------------------

A video demonstration is available at \url{https://t.ly/QTSdp}, and tool artifacts and documentation can be found at \url{https://t.ly/7PSFv}.

%-----------------------------------------------------
\section{Conclusions and Future Work}
%-----------------------------------------------------

This paper introduces ESBMC-Python, a novel tool to detect errors in Python programs. It builds a front-end for the ESBMC framework that translates Python expressions into symbols, which are converted into a GOTO program, symbolically executed, and checked for satisfiability. To the best of our knowledge, there is only one tool \cite{shu2019model} to formally verify Python code with model checking, which highlights the importance of our work. However, a direct comparison was not possible due to code and result unavailability.

To evaluate ESBMC-Python's effectiveness, we conducted experiments using a test suite. On average, simple programs could be verified in $31.56$ms. Moreover, such a suite can also serve as a reference for evaluating other Python verification tools. Additionally, ESBMC-Python was able to identify a real bug in the specification of the Ethereum blockchain consensus protocol.

In future work, we intend to extend ESBMC-Python's capabilities to include more features. Additionally, we aim to enhance our type inference algorithm to deal with more complex program flows. The potential of large language models will also be explored to address type inference for complex expressions and execution paths. Finally, we plan to develop operational models for verifying AI libraries.

\section*{Acknowledgments}
This project was supported by the Ethereum Foundation under Grant FY22-0751.

%Bibliography
% \clearpage
\bibliographystyle{unsrt}  
\bibliography{references}

\end{document}